\documentclass[prl,floatfix,superscriptaddress,nofootinbib,noshowpacs,showkeys,preprintnumbers]{revtex4}

\usepackage[dvips]{epsfig}
\usepackage{graphicx}

\begin{document}

\title{Observation of narrow baryon resonance in $pK^0_s$ mode
in $pA$-interactions at $70\ GeV/c$ with SVD-2 setup.}

\author{A.~Kubarovsky, V.~Popov \\ (for the SVD Collaboration)\\
D.V. Skobeltsyn Institute of Nuclear
Physics, Lomonosov Moscow State University, \\
1/2 Vorobyevy gory, Moscow, 119992 Russia\\
E-mail: alex\_k@hep.sinp.msu.ru }


\begin{abstract}
\noindent We report on the SVD-2 experiment data analysis aimed to
search for an exotic baryon state, the $\Theta^+$-baryon, in a
$pK^0_s$  decay mode with IHEP U-70 accelerator proton beam at
$70~GeV/c$. A resonant structure with $M=1526\pm3(stat.)\pm
3(syst.)~MeV/c^2$ and $\Gamma < 24~MeV/c^2$ was found in the $pK^0_s$
invariant mass spectrum, with the statistical significance of this peak
estimated as $5.6~\sigma$.
\end{abstract}

\maketitle

\section{SVD-2 experiment}
The present analysis was made in a framework of SVD-2 experiment, the
main goal of which is a study of the charm hadroproduction at the
near-threshold  energy\cite{svd0,svd1,svd2}. SVD-2 setup consists
of high-precision microstrip vertex detector (beam telescope, active
target and tracking detector), large aperture magnetic spectrometer,
multicell threshold Cherenkov counter and Cerenkov full absorption
lead glass gamma detector.

The primary vertex position determination procedure was based on
well-known "tear-down" approach \cite{tear1, tear2}.
Only events with a well reconstructed primary vertex were selected.
After excluding the tracks that belongs to primary vertex, the
secondary vertex position was determined by finding V0-decay
downstream the primary vertex. The primary vertex resolution was
estimated as $70-120~\mu m$ for Z-coordinate and $8-12~\mu m$ for
X(Y)-coordinates. For the two-tracks secondary vertices  ($K^0_s,\
\Lambda$) those values were $250~\mu m$ and $15~\mu m$ respectively.
The impact parameter resolution for $3-5\ GeV$ momentum tracks is
about  $12~\mu m$. The angular acceptance of the vertex detector
averages to $\pm 250\ mrad$.

\begin{figure}[h]
\begin{center}
\begin{minipage}[b]{0.49\textwidth}
\centering
\includegraphics[width=0.8\textwidth]{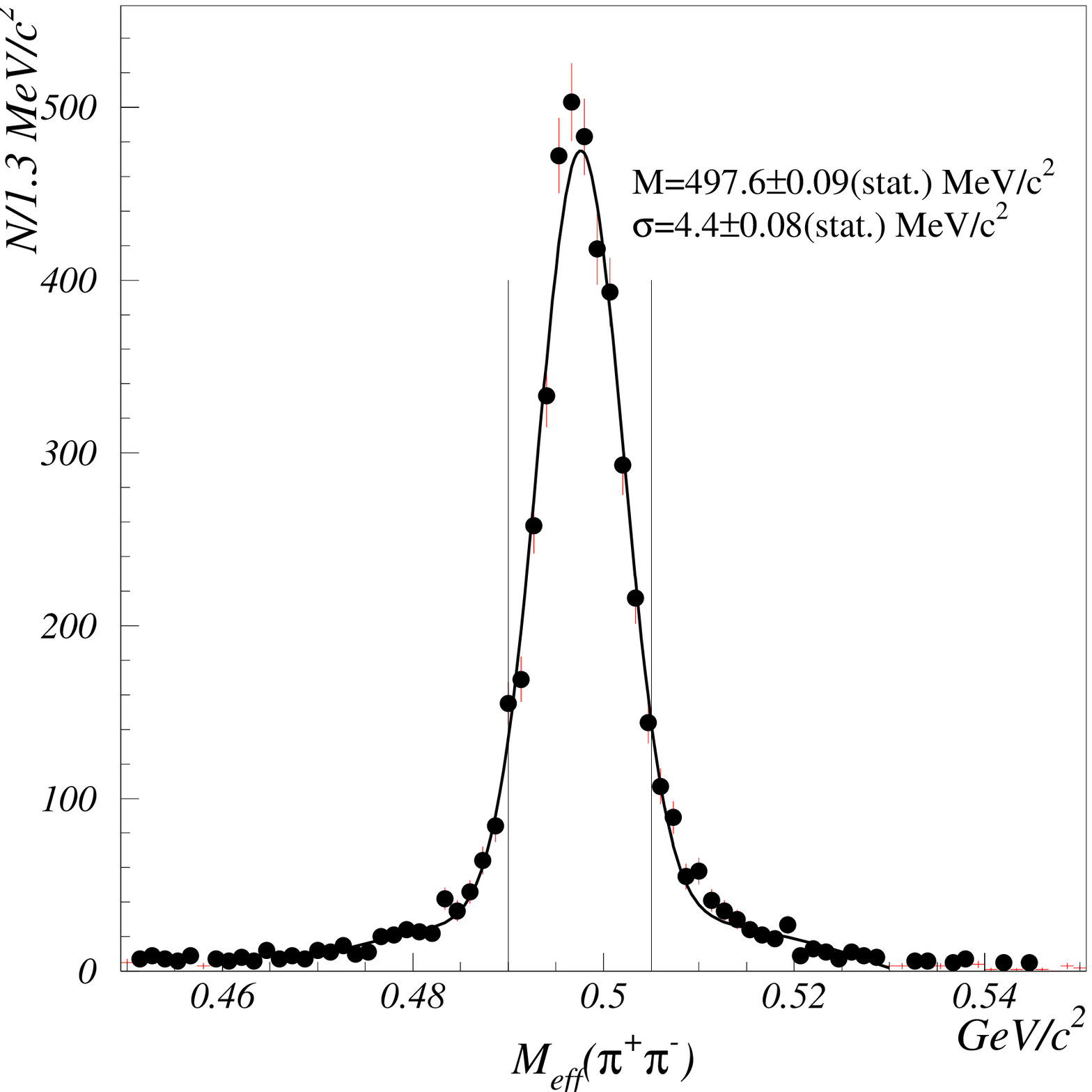}
\end{minipage}
\begin{minipage}[b]{0.49\textwidth}
\centering
\includegraphics[width=0.8\textwidth]{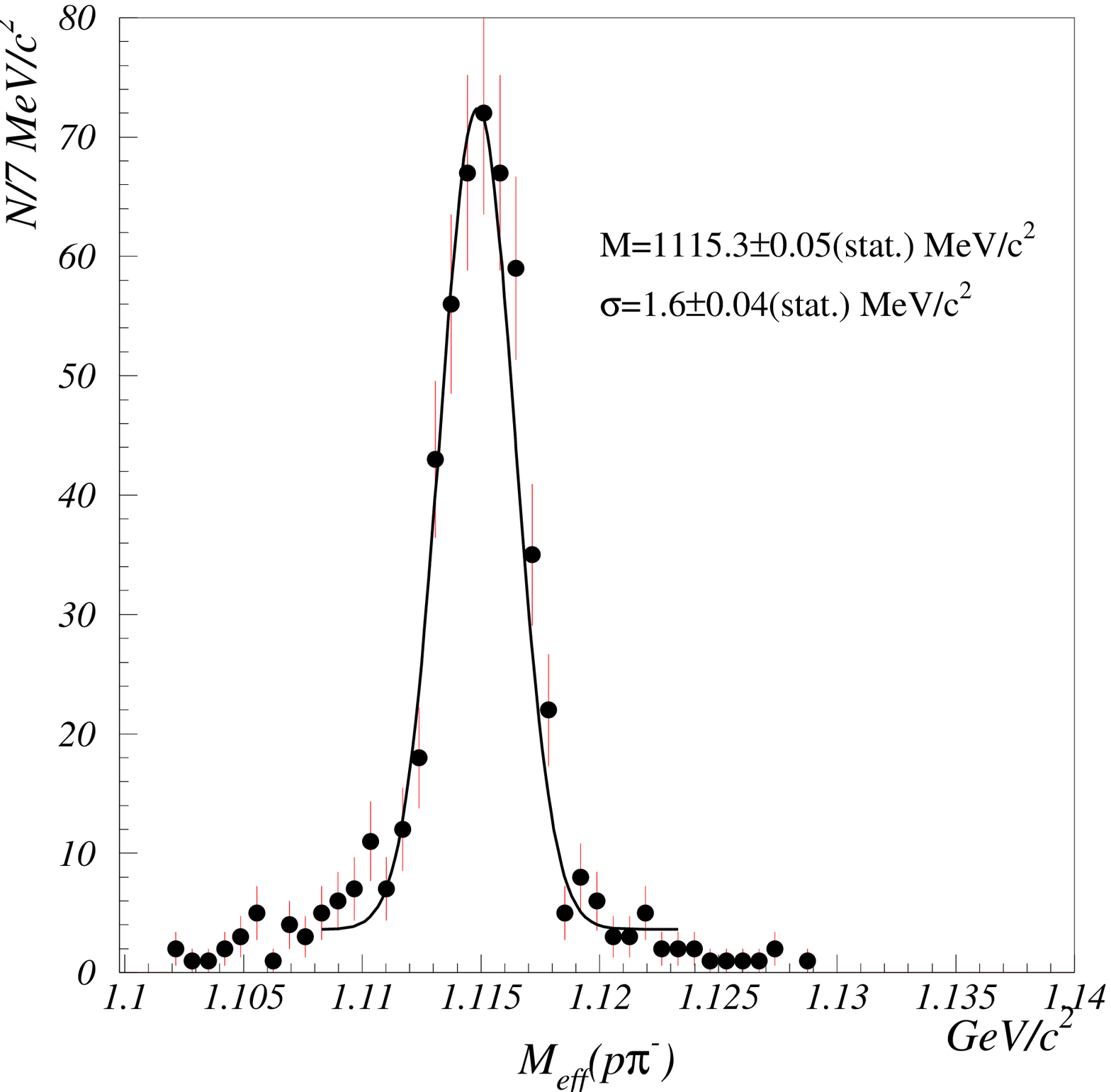}
\end{minipage}
\caption[*]{Left: the $(\pi^+\pi^-)$ invariant mass spectrum. A window
corresponding to $\pm2\sigma$ is shown by the vertical lines.
Right: the $(p \pi^-)$ invariant mass spectrum.}
\label{k0_lambda_mass}
\end{center}
\end{figure}

The SVD-2 setup permits obtaining the high effective
mass resolution of $\sigma = 4.4~MeV/c^2$ for $K^0_s$ and $1.6~MeV/c^2$
for $\Lambda^0$ masses (see fig.\ref{k0_lambda_mass}). The
momentum resolution for the track with 15 measured hits is
$(0.5\div1.0)\%$ in the $(4\div20)\ GeV/c$ momentum range. The
angular measurement error was estimated to be
$0.2\div0.3\ mrad$. The angular acceptance of spectrometer
averages to $\pm200\ mrad$ for horizontal and $\pm150\ mrad$ for
vertical coordinates.

\begin{figure}[h]
\begin{center}
\includegraphics[width=0.4\textwidth]{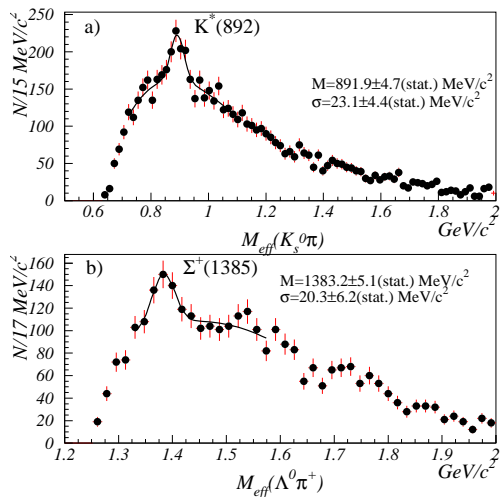}
\caption[*]{a) The combined $(\pi^+K^0_s)$ and $(\pi^-K^0_s)$ invariant
mass spectrum. b) $(\Lambda^0 \pi^+)$ invariant mass spectrum.}
\label{sigma_ks_mass}
\end{center}
\end{figure}

The combined $(\pi^+K^0_s)$ and $(\pi^-K^0_s)$ invariant mass spectrum
is shown on fig.\ref{sigma_ks_mass}a. The $K^*(892)$ peak is clearly
seen on the distribution. Fig.\ref{sigma_ks_mass}b shows $(\Lambda^0 \pi^+)$
invariant mass spectrum. $\Sigma^+(1385)$ peak is clearly seen. The masses of
observed $K^0_s$, $\Lambda^0$ and also masses and widths of $K^*(892)$
and $\Sigma^+(1385)$ are consistent with their PDG values\cite{pdg}.
\section{$\Theta^+$-baryon searches}
Exotic baryons with 5-quarks content (pentaquarks) and their
properties have been predicted by Diakonov, Petrov, and Polyakov
\cite{Diakonov} in the framework of the chiral soliton model, although
such 5-quarks structures were proposed years ago
\cite{jaffe1,pras}. The lightest member of the pentaquarks
antidecuplet, $\Theta^+$-baryon, has
positive strangeness, mass $M \sim 1530~MeV/c^2$,~$\Gamma \le
15~MeV/c^2$, spin $1/2$ and even parity.

Experimental evidence for $\Theta^+$-baryon with positive strangeness
came recently from several experimental groups (LEPS\cite{Nakano},
DIANA-ITEP \cite{Dolgolenko,itep}, CLAS \cite{Ken,vpk,clas2}, SAPHIR
\cite{SAPHIR}). In those experiments $\Theta^+$-baryon was observed in
the $nK^+$ or $pK^0_s$ invariant mass spectra with the mass near
$1540~MeV/c^2$. More recently HERMES collaboration observed narrow
baryon state at the mass of $1528\ MeV/c^2$ in quasi-real
photoproduction\cite{hermes}. Also ZEUS collaboration \cite{zeus}
reported an evidence of the exotic baryon in $pK^0_s$-channel with the
mass of $1527\ MeV/c^2$.

\section {Events selection and $pK^0_s$-spectrum analysis.}
We were searching for the $\Theta^+$-baryon in the reaction
$pN\rightarrow \Theta^+ + X$, \ $\Theta^+ \rightarrow pK^0_s$, \
$K^0_s \rightarrow \pi^+\pi^-$.
The data pre-selected at the search for the charm production were used.
It consisted of events with V0-candidates decayed within the vertex
detector: decay length $\le 35\ mm$ with a mean of
$20\ mm$. The following criteria were then applied:
\begin{itemize}
\item
Primary charged particles multiplicity $\le 5\/$. Minimizes the
combinatorial background, reduces the probability
of appearance of the events with rescattering on nuclei and
background of $K^0_s$-mesons produced in the central rapidity
region.
\item
A presence of proton as well-measured primary positive track with
a momentum of $4\ GeV/c \le P_p \le 21\ GeV/c$ leaving no hit in
Cherenkov counter.
\item
$490\ MeV/c^2 \le M_{\pi^+\pi^-} \le 505\ MeV/c^2$ to select
  well-identified $K^0_s$.
\item
$cos(\alpha) \ge 0$, where $\alpha$ is angle of flight of
$pK^0_s$-system in the center mass system of the beam proton and the
target nucleon(beam proton fragmentation region).
\item
$P_{K^0_s}\le~P_p$ kinematical cut \cite{levtch}: effectively destroys
most of the decays of $\Sigma^{*+}$-resonances with high masses while
cutting only $10\%$ of $\Theta^+$-peak events.
\end{itemize}

\begin{figure}[h]
\centering
\includegraphics[width=0.5\textwidth]{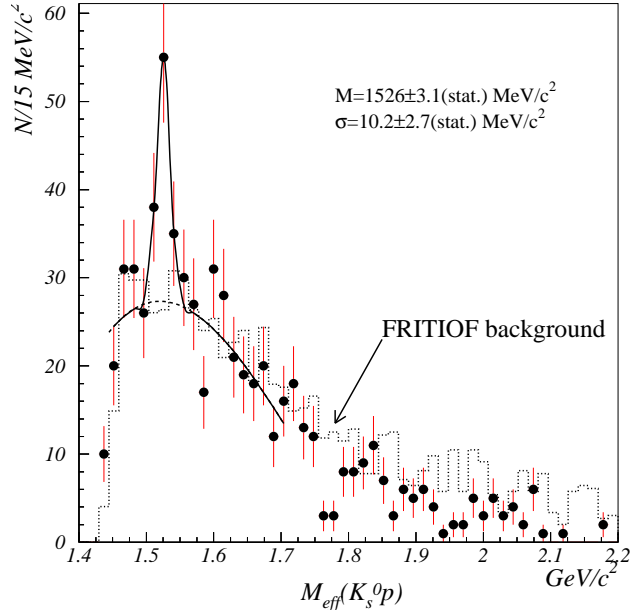}
\caption[*]{The $(pK^0_s)$ invariant mass spectrum in the reaction
$pA\rightarrow pK^0_s+X$. Dashed line: background obtained from FRITIOF
simulations.}
\label{theta}
\end{figure}

Resulting distribution is shown on fig. \ref{theta}.  The distribution was
fitted by Gaussian function and fourth-order polynomial
background. There are 50 events in the peak over 78
background events. The statistical significance for the fit inside a
$45\ MeV/c^2$ mass window was calculated as
$N_{P}/\sqrt{N_{B}}$, where $N_{B}$ is the number of counts in the
background fit under the peak and $N_{P}$ is the number of counts in
the peak. We estimate the significance to be of $5.6~\sigma$. It is
impossible to determine the strangeness of this state in such an
inclusive reaction, however there are no reported
$\Sigma^{*+}$-resonances in $1500\div1550~MeV/c^2$ mass area, so we
interpret observed state as recently reported $\Theta^+$-baryon with
a positive strangeness.

\begin{figure}[h]
\centering
\includegraphics[width=0.35\textwidth,clip,trim=10 10 10 10]{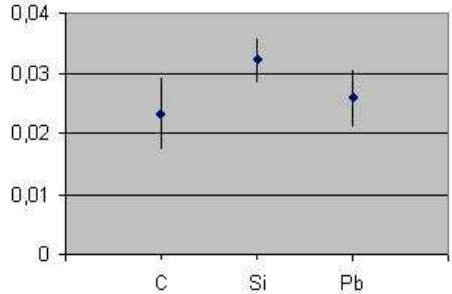}
\caption[*]{The ratio of $\Theta^+$ events to the total of $K^0_s$
  events for the different target materials}
\label{adepend}
\end{figure}

The A-dependence analysis in the observed peak area showed no
difference (within measuring errors) from the $A^{0.7}$
dependence for background inelastic events (fig. \ref{adepend}).

In a summary, the inclusive reaction $p A \rightarrow pK^0_s + X$ was
studied at IHEP accelerator with proton energies at $70\ GeV$ using
SVD-2 detector. With several cuts applied a narrow baryon resonance
was observed with mass $M=1526\pm 3(stat.)\pm 3(syst.)\ MeV/c^2$ and
$\Gamma < 24\ MeV/c^2$. The width of this state is close to SVD-2
experimental resolution for $pK^0_s$-system and its mass and width are
consistent with recently reported $\Theta^+$-resonance\cite{Nakano,
Dolgolenko,Ken,vpk,SAPHIR,itep,clas2,hermes}, which was predicted as
an exotic pentaquark ($uudd\bar{s}$) baryon state.

\newpage

\section*{Acknowledgements} We thank HSQCD'04 Organizing Committee and
personally V.~Kim for providing the excellent, warm and stimulating
atmosphere during the Conference and for the financial support.
We are grateful to V.~Kubarovsky, B.~Levchenko and N.~Zotov for useful
comments and suggestions.

\end{document}